\documentclass[copyright,creativecommons]{eptcs}

\usepackage{iftex}

\ifpdf
  \usepackage{underscore}         % Only needed if you use pdflatex.
  \usepackage[T1]{fontenc}        % Recommended with pdflatex
\else
  \usepackage{breakurl}           % Not needed if you use pdflatex only.
\fi

\usepackage[linesnumbered]{algorithm2e}
\usepackage{algorithmic}
\usepackage{graphicx}
\usepackage{caption}
\usepackage{subcaption}
\usepackage{textcomp}
\usepackage{xcolor}
\usepackage{listings}
\usepackage{xspace}
\usepackage{multirow}
\usepackage{pifont}
\usepackage{array}
\usepackage{amsmath}
\usepackage{wrapfig}

\newcommand{\removelatexerror}{\let\@latex@error\@gobble}

\definecolor{lbcolor}{rgb}{0.875,0.875,0.875}
\newcommand{\rlibm}{\textsc{RLibm}\xspace}

\newcommand{\eg}{\emph{e.g.}\xspace}
\newcommand{\ie}{\emph{i.e.}\xspace}

\title{Fast Trigonometric Functions using the RLIBM Approach}
\author{Sehyeok Park \institute{Rutgers University\\ New Brunswick,
    NJ, USA} \email{sp2044@cs.rutgers.edu} \and Santosh Nagarakatte
  \institute{Rutgers University\\ New Brunswick, NJ, USA}
  \email{santosh.nagarakatte@cs.rutgers.edu} } \def\titlerunning{Fast Trigonometric Functions using the RLIBM Approach}
\def\authorrunning{S. Park \& S. Nagarakatte}
\begin{document}
\maketitle

\begin{abstract}
This paper describes our experience developing polynomial
approximations for trigonometric functions that produce correctly
rounded results for multiple representations and rounding modes using
the \rlibm approach. A key challenge with trigonometric functions
concerns range reduction with $\pi$, which reduces a given input in
the domain of a 32-bit float to a small domain.  Any rounding error in
the value of $\pi$ is amplified during range reduction, which can
result in wrong results.
We describe our experience implementing fast range reduction
techniques that maintain a large number of bits of $\pi$ both with
floating-point and integer computations.
The resulting implementations for trigonometric functions are fast and
produce correctly rounded results for all inputs for multiple
representations up to 32-bits with a single implementation.
\end{abstract}

\section{\textbf{Introduction}}
Scientific computing extensively uses elementary functions~(\eg,
$e^x$) provided by math libraries. Producing correctly rounded results
for all inputs for elementary functions is challenging (\ie,
table-maker's dilemma~\cite{Kahan:tablemaker:online:2004}). Hence,
mainstream math libraries do not produce correct results for all
inputs. The lack of correctly rounded math libraries can make the
application non-portable and can cause reproducibility issues. An
application can produce totally different results on two different
machines.

%Correctly rounded math libraries can avoid this problem.

A common approach to develop math libraries is to generate minimax
polynomial
approximations~\cite{Chevillard:sollya:icms:2010,Markstein:2000:IA64},
which minimize the maximum error across all inputs with respect to the
real value. Eventually, the real coefficients of the generated
polynomial are rounded to a hardware supported FP
representation~\cite{Brisebarre:maceffi:toms:2006,Brisebarre:epl:arith:2007}. By
using polynomial approximations with sufficiently large degrees, such
minimax approaches can generate correctly rounded elementary functions
for a single representation~\cite{Daramy:crlibm:spie:2003}.

As an alternative, our \rlibm
project~\cite{lim:rlibmall:popl:2022,lim:rlibm:popl:2021,lim:rlibm32:pldi:2021,
  aanjaneya:rlibm-prog:pldi:2022} makes a case for approximating the
correctly rounded result directly rather than the real value of an
elementary function. The \rlibm project uses the MPFR high-precision
math library~\cite{Fousse:toms:2007:mpfr} as the oracle and focuses on
generating efficient implementations given the oracle result.
Once the oracle correctly rounded result for a given input is known,
the main insight in the \rlibm project is that there is an interval of
real values around the correctly rounded result such that producing
any real value in this interval rounds to the correctly rounded
result. %
With the \rlibm approach, the freedom available to the polynomial
generation step is larger than the freedom available with minimax
methods (\ie, it is 1 units in the last place, ULP, for all
inputs). In contrast, minimax methods have significantly smaller
freedom when the real-valued output of an elementary function is very
close to the rounding boundary (\eg, midpoint of two FP values with
round to nearest
modes)~\cite{Lefevre:worstcase:arith:2001,Abraham:fastcorrect:toms:1991}.
%
%% In such cases, the freedom with the minimax methods is
%% the distance between the real value and the rounding boundary, which
%% is significantly smaller than 1 ULP.
%
Using the interval around the correctly rounded result, the \rlibm
project structures the task of generating a polynomial of degree~$d$
that produces correctly rounded results for all inputs as a linear
program (LP) (\ie, a system of linear inequalities). It uses an LP
solver to identify the
coefficients~\cite{aanjaneya:rlibm-prog:pldi:2022,aanjaneya:max-consensus:pldi:2024}.

With the increased use of custom
formats~\cite{nvidia:tensorfloat:online:2020,Wang:tpu:online:2019},
the \rlibm project proposes a novel idea to generate a single
polynomial approximation that produces correctly rounded results for
multiple representations and rounding
modes~\cite{lim:rlibmall:popl:2022}, which can serve as a reference
library.  To generate correctly rounded results for FP representations
with up to $n$-bits that have $E$-bits for the exponent, the key
insight is to generate a polynomial approximation that produces
correctly rounded results for a representation with $(n+2)$-bits with
the \emph{round-to-odd} rounding
mode~\cite{lim:rlibmall:popl:2022}. When such a result is double
rounded to the target representation, it produces correct results for
representations with $k$ bits, where $ E+2 \leq k \leq n$, and for all
standard rounding modes.

We have been trying to generate fast correctly rounded trigonometric
functions with the \rlibm approach for a few years. We were not
successful earlier because we did not know how to perform both correct
and efficient range reduction for these functions. The excellent
reference on elementary functions by
Muller~\cite{Muller:elemfunc:book:2016} describes algorithms for range
reduction for trigonometric functions. However, na\"{\i}vely implementing
these algorithms can slow down resulting implementations by
2-3$\times$ when compared to final fast implementations that we
describe in this paper.
Further, it is necessary to compute with a large number of digits of
$\pi$ efficiently. When we maintained an insufficient number of bits
of $\pi$, numerical errors in range reduction performed with double
precision often resulted in conflicting constraints in the linear
program generated with the \rlibm approach.

The first step for generating any correctly rounded elementary
function is range reduction, which reduces an input in the domain of
floating-point (FP) to a small domain where a polynomial approximation
is feasible.  A key challenge with range reduction for trigonometric
functions is that one would need to maintain a large number of bits of
$\frac{1}{\pi}$. Given an input $x$, one can perform range reduction
as $x' = |x| - k\pi$ to produce the reduced input $x'\in [0, \pi)$,
  where $k=\lfloor \frac{|x|}{\pi} \rfloor$. Alternatively, one can
  also perform a symmetric range reduction $x-k\pi$ using
  $k=[\frac{x}{\pi}]$, where $[\frac{x}{\pi}]$ computes the nearest
  integer when $\frac{x}{\pi}$ is rounded, to obtain $x' \in
  [-\frac{\pi}{2}, \frac{\pi}{2}]$. If such range reduction is applied
  using just 64-bits of $\frac{1}{\pi}$, the reduced input will have
  no accurate bits when $x$ is a large number.
%
%% Another challenge to address when one does maintain a large number of
%% bits for $\pi$ concerns the efficiency of range reduction.

This paper describes our experience exploring efficient range
reduction techniques for trigonometric functions, which is performed
using a combination of FP and integer operations, for generating a
single polynomial approximation that produces correct results for all
inputs with multiple representations and rounding modes. Although the
core algorithms for range reduction of trigonometric functions are
well-known~\cite{payne-haneck:radian-reduction:signum:1983,
  Muller:elemfunc:book:2016,Ng92argumentreduction, daumas1996modular,
  drmrr:2006,
  Daramy:crlibm:spie:2003,Daramy:crlibm:doc,762822,Lefevre:toward:tc:1998,Cody:book:1980},
the challenge is in efficiently implementing them with sufficient
accuracy to produce correctly rounded results for all inputs. We
observe that efficient range reduction is as important or more
important than low-degree polynomial approximations for trigonometric
functions.  We describe the algorithms and evaluate the performance
implications of various range reduction techniques. We believe this
exposition will help future implementers of math libraries.
Using these range reduction techniques, we have implemented correctly
rounded trigonometric functions where a single implementation produces
correctly rounded results for multiple representations from 10-bits to
32-bits for all the standard rounding modes. It is faster or similar
in performance compared to mainstream libraries and other correctly
rounded libraries. Our integer-based range reduction improves the
performance by 19\% when compared to our FP-based strategy. \emph{In
summary, the efficient range reduction methods described in this paper
enabled us to generate correct and fast trigonometric functions after
multiple years of trying to use the \rlibm approach for these
functions}.

\section{\textbf{Background on the \rlibm Approach}}
The \rlibm approach~\cite{lim:rlibm:popl:2021,
  lim:rlibm32:pldi:2021,lim:rlibmall:popl:2022,aanjaneya:rlibm-prog:pldi:2022}
assumes the existence of an oracle that provides correctly rounded
results (\eg, the MPFR library) and focuses on generating efficient
implementations. Given the oracle result, the \rlibm project makes a
case for approximating the correctly rounded result rather than the
real value of an elementary function.
\begin{figure}
  \centering{\includegraphics[width=0.8\textwidth]{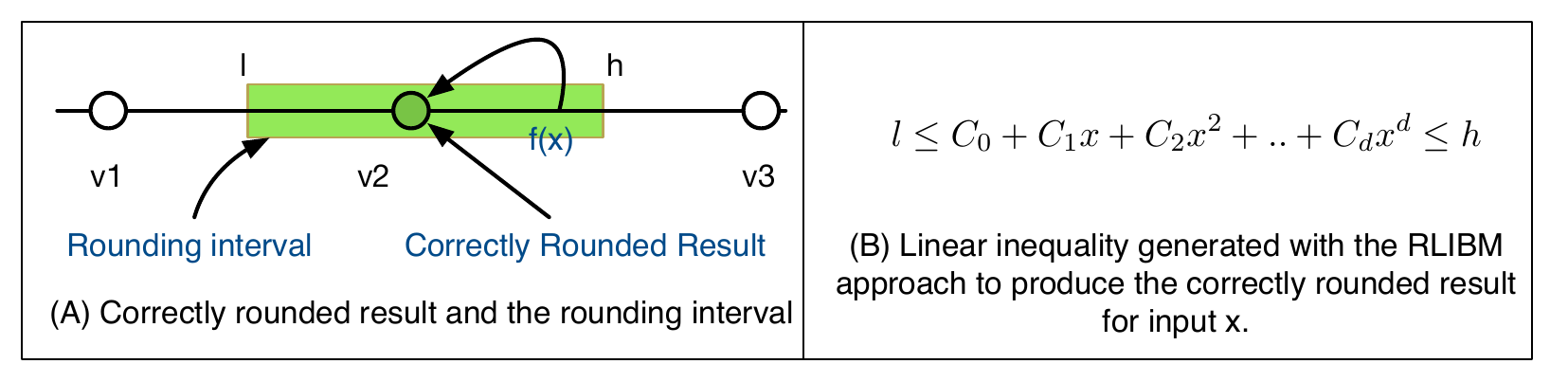}}
  \caption{\small (A) The rounding interval of a correctly rounded
    result v2. (B) The linear inequality generated with the \rlibm
    approach to produce the correctly rounded result for an input $x$
    given the rounding interval $[l, h]$.}
  \label{fig:background}
\end{figure}

Typically, the implementation of a correctly rounded function for a
32-bit float representation uses the 64-bit double precision
representation internally.  After a correctly rounded result for a
32-bit float input is available using an oracle, the \rlibm project
observes that there is an interval of values in the double precision
representation around the correctly rounded result such that any value
in that interval rounds to the correctly rounded value (see
Figure~\ref{fig:background}(A)). This interval is called the rounding
interval. The rounding interval is represented as $[l, h]$, where $l$
is the lower bound and $h$ is the upper bound.  When the goal is to
generate a polynomial of degree~$d$ with $d+1$ terms, the rounding
interval implies a linear constraint on the result of the polynomial
evaluation for a given input $x$ as shown in
Figure~\ref{fig:background}(B). A 32-bit representation has $2^{32}$
inputs and the \rlibm project generates a system of linear
inequalities corresponding to the 32-bit inputs and their respective
rounding intervals.  Now, the task of generating a correctly rounded
function boils down to the task of identifying the coefficients (\ie,
$C_i$'s) of a polynomial that satisfies these inequalities. Hence, the
\rlibm project frames the problem of generating correct elementary
functions as a linear program and uses an LP solver to solve it.

Range reduction techniques that reduce the original input from the
domain of a 32-bit float representation (\ie, $[2^{-149}, 2^{128})$)
  to a small domain (\eg, $[-1, 1]$) are crucial to generate
  approximations. The original input $x$ is range reduced to $x'$.
The polynomial approximation computes the result for $x'$. The result
is output compensated to produce the final output for $x$.
Both range reduction and output compensation are performed in the
double precision representation and can have numerical errors.
To account for the numerical errors, the \rlibm project deduces
intervals for the reduced domain such that the polynomial evaluation
over the reduced input produces correct results for the original
inputs. The \rlibm project uses the inverse of the output compensation
function to infer the reduced rounding intervals.  Finally, a system
of linear inequalities corresponding to the reduced inputs and the
reduced rounding intervals are solved using an LP solver iteratively
to identify coefficients of a polynomial of degree~$d$.

\begin{figure}
  \centering{\includegraphics[width=0.98\textwidth]{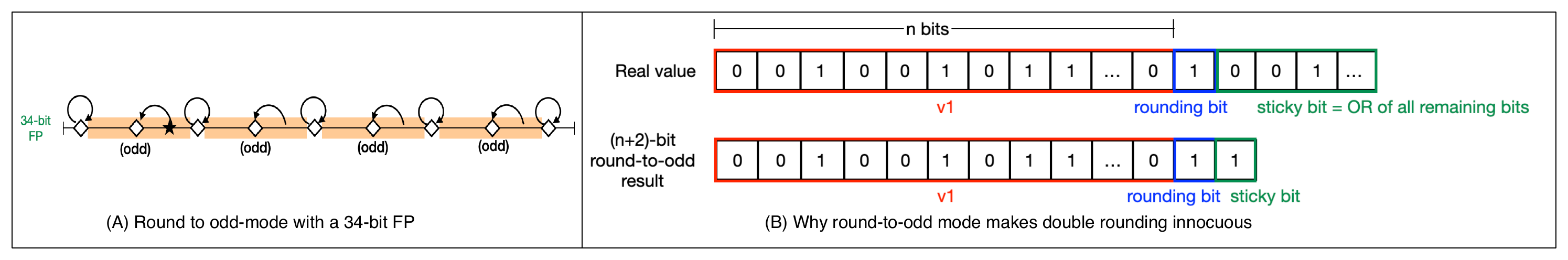}}
  \caption{\small (A) Illustration of round-to-odd rounding mode with
    a 34-bit FP representation. (B) Intuition on why double rounding
    is innocuous with the round-to-odd mode. Rounding the round-to-odd
    result to a target representation has the same truncated value,
    rounding bit, and sticky bit as rounding that real value directly
    to the given target representation.}
  \label{fig:rno}
\end{figure}

\textbf{Multiple representations and rounding modes}. The \rlibm
approach described till now produces correctly rounded results for all
inputs for a single representation and a single rounding mode. There
are four rounding modes in the standard. Further, many new
representations that vary either the dynamic range or the precision
are being proposed (\eg, bfloat16, tensorfloat32, FP8). A single
implementation that provides correct results for these representations
is appealing as a reference library. One approach to generate a single
approximation for multiple representations is to use a correctly
rounded function designed for a higher precision representation. Then,
round the result to the target representation.  However, it produces
wrong results for some inputs because of double rounding errors. The
first rounding happens when the real value is rounded to produce the
correctly rounded result of a higher precision representation. The
second rounding happens when that result is rounded to the target
representation.

The \rlibm project has proposed a method to generate a single
polynomial approximation that produces correctly rounded results for
all FP representations up to
$n$-bits~\cite{lim:rlibmall:popl:2022}. The key idea is to generate a
polynomial for the ($n$+2)-bit representation (which has 2 additional
precision bits compared to the $n$-bit representation) using the
\emph{round-to-odd} rounding mode. The round-to-odd result with two
additional precision bits retains all necessary information to produce
the correct results for any representation up to $n$-bits.
In the round-to-odd mode~\cite{boldo:rno:imacs:2005, goldberg:fp}, if
a real value is exactly representable by the target representation, we
represent it with that value. Otherwise, the value is rounded to an
adjacent floating point value whose bit-pattern is odd when
interpreted as an integer. Figure~\ref{fig:rno}(A) illustrates the
round-to-odd mode.
To understand why double rounding with the round-to-odd mode produces
correct results, we have to first understand how rounding
works. Typically, we need three pieces of information when rounding a
real value to an $n$-bit FP representation: (1) the first $n$-bits of
the real value in the binary representation, (2) the $(n+1)^{th}$-bit
known as the rounding bit, and (3) the result of the bitwise OR of all
the remaining bits, known as the sticky bit.
When we round a real value to a ($n$+2)-bit representation using the
round-to-odd mode, the round-to-odd result precisely maintains the
same three pieces of information as rounding the real value directly
to the target representation of a lower bitwidth as shown in
Figure~\ref{fig:rno}(B). Hence, subsequent rounding of the
round-to-odd result to a representation with less than or equal to
$n$-bits produces the correct result. This ability to produce
correctly rounded results for multiple representations and rounding
modes with a single polynomial approximation is feasible because the
\rlibm project directly approximates the correctly rounded result
rather than the real value.

\iffalse
In contrast to the \rlibm approach, minimax methods do not take
the rounding mode and rounding into account and hence, they do not
produce correctly rounded results for multiple representations and
rounding modes with a single polynomial approximation. From the
perspective of building reusable and portable libraries, developing a
single polynomial approximation that produces correctly rounded
results for multiple representations and rounding modes is attractive.
\fi

\section{Fast Range Reduction with Sufficient Accuracy}
\label{sec:approach:rangereduction}

We describe range reduction techniques and implementation choices that
finally enabled us to build fast and correct trigonometric functions
with the \rlibm approach. During range reduction, the original input
$x$ in the domain of a 32-bit float is mapped to a reduced input
$x'$. This process involves $\pi$ (\ie, $x' = x - k \pi/2^t$). For
large values of $x$, we need to maintain a large number of bits of
$\frac{2^{t}}{\pi}$.  We then need to perform computation with them
efficiently to identify $k = [\frac{2^t x}{\pi}]$ and the reduced
input. If we just maintain 64 precision bits, then the reduced input
is either 0 (for large inputs) or significantly wrong (for other
inputs). This range reduction is extremely challenging when the goal
is to produce correctly rounded results for all inputs while providing
good performance! Existing libraries make trade-offs either with
performance (\eg, Payne-Hanek
implementations~\cite{payne-haneck:radian-reduction:signum:1983}) or
correctness (\eg, GCC's libm).
Trigonometric functions are periodic functions with a period of
$2\pi$. Hence, the range reduction produces a reduced input $x' \in
[-\frac{\pi}{2^{t+1}}, \frac{\pi}{2^{t+1}}]$ given the original input
$x \in [-\infty, \infty]$ using periodicity, where $t$ is an integer
greater than or equal to $0$. Effectively,

\begin{equation}
    x =  x' + k \pi/2^t \\ \nonumber
\end{equation}
To perform the above range reduction, the key task involves computing
an integer $k = [\frac{x}{\frac{\pi}{2^t}}] = [ \frac{2^t x}{\pi}
]$. Using $k$, one can compute a reduced input in
$[-\frac{\pi}{2^{t+1}}, \frac{\pi}{2^{t+1}}]$ as shown below.

\begin{equation}
  \label{eqn-reduced}
   r = \frac{2^tx}{\pi} - k, \  \  \ \  x' = x - k \frac{\pi}{2^t} = \frac{\pi}{2^t} \left(\frac{2^t
     x}{\pi} - k\right) = \frac{\pi}{2^t}r \\
\end{equation}

These identities hold when the computation is performed with real
numbers. For this range reduction, we need to perform computations
involving $\frac{2^t}{\pi}$.  Doing so requires maintaining a large
number of bits of $\frac{2^{t}}{\pi}$. Without maintaining sufficient
digits of $\frac{2^t}{\pi}$, the resulting reduced input will have very
few accurate bits. This is because most of the leading bits will be
canceled when $x$ is close to a multiple of $\frac{\pi}{2^t}$ and the
remaining bits will be influenced by rounding. This problem is
extremely severe when $x$ is a large number. If we maintain
$\frac{256}{\pi}$ using only 64 bits, assuming $t=8$, then the reduced
input will be 0 for all inputs when $2^{80} \le x$. This is because
all the 64 precision bits of $\frac{256}{\pi}$ would contribute to
$k$, leaving no bits available for the fractional piece
$r=\frac{256x}{\pi} - k$.

\textbf{Range reduction for $\mathbf{sin(x)}$}.
As described above, the main objective of range reduction for $sin(x)$
is to transform a given input $x \in [-\infty, \infty]$ into a reduced
input $x' \in [-\frac{\pi}{2^{t+1}}, \frac{\pi}{2^{t+1}}]$. Leveraging
the formula $x = k\frac{\pi}{2^t} + x'$ and the trigonometric property
$sin(a+b) = sin(a)cos(b) + cos(a)sin(b)$, we can compute $sin(x)$ as
follows. 

\begin{align}
  \label{sin-range-reduction-initial}
  sin(x) =& sin\left(\frac{k\pi}{2^t} + x'\right) = sin\left(\frac{k\pi}{2^t}\right)cos(x') + cos\left(\frac{k\pi}{2^t}\right)sin(x') 
\end{align}

Equation~\ref{sin-range-reduction-initial} reduces the task of
generating a polynomial approximation of $sin(x)$ for an input $x \in
[-\infty, \infty]$ to generating approximations of $sin(x')$ and
$cos(x')$ for the reduced input $x' \in
[-\frac{\pi}{2^{t+1}}, \frac{\pi}{2^{t+1}}]$.
Once we choose a value for $t$ (\eg, $t=8$ and we generate a reduced
input in $[-\frac{\pi}{512}, \frac{\pi}{512}]$), the output
compensation formula in Equation~~\ref{sin-range-reduction-initial}
requires us to precompute the tables for $sin(\frac{k\pi}{2^t})$ and
$cos(\frac{k\pi}{2^t})$ for all possible values of $k$.
One issue in precomputing these tables is that the set of possible
values of $k$ can be very large. We leverage the periodicity and
symmetry of $sin(x)$ and $cos(x)$ to reduce the number of necessary
precomputed values.

Specifically, $sin(x) = sin(x - 2m\pi)$ and $cos(x) = cos(x - 2m\pi)$
for any integer $m$.  Applying this property to
$sin(\frac{k\pi}{2^t})$ and $cos(\frac{k\pi}{2^t})$ in
Equation~\ref{sin-range-reduction-initial} leads to a significant
reduction in table sizes.

\begin{equation}
  \label{sin-m-k}
  m = \lfloor \frac{k}{2^{t+1}} \rfloor, \  \  k' = k - 2^{t+1} m 
\end{equation}
\begin{align}
  \label{sin-range-reduction-table1}
  sin\left(\frac{k\pi}{2^t}\right) =&  sin\left(\frac{k\pi}{2^t} - 2m\pi\right) = sin\left(\frac{k - 2^{t+1} m}{2^t}\pi\right) = sin\left(\frac{k'\pi}{2^t}\right)
\end{align}
\begin{align}
    \label{sin-range-reduction-table2}
    cos\left(\frac{k\pi}{2^t}\right) =& cos\left(\frac{k\pi}{2^t} - 2m\pi\right) =  cos\left(\frac{k - 2^{t+1} m}{2^t}\pi\right) = cos\left(\frac{k'\pi}{2^t}\right)
\end{align}

Applying Equation~\ref{sin-range-reduction-table1} and
Equation~\ref{sin-range-reduction-table2} to
Equation~\ref{sin-range-reduction-initial}, we get 

\begin{equation}
  \label{sin-almost-final}
  sin(x) = sin\left(\frac{k'\pi}{2^t}\right)cos(x') + cos\left(\frac{k'\pi}{2^t}\right)sin(x')
\end{equation}

A total of $2^{t+2}$ precomputed values are
necessary at this point ($2^{t+1}$ each for $sin(\frac{k'\pi}{2^t})$
for $cos(\frac{k'\pi}{2^t})$), since $k' \in [0, 2^{t+1})$. If $t=8$,
then would need a total of $1024$ precomputed values. We further
reduce the number of required table entries to $512$ by exploiting the
property that $cos(\frac{k'\pi}{2^t}) = sin(\frac{k'\pi}{2^t}
+ \frac{\pi}{2}) = sin(\frac{k'+2^{t-1}}{2^t}\pi)$. Using the same
property as before, one can conclude that
$sin(\frac{k'+2^{t-1}}{2^t}\pi) =
sin(\frac{k'+2^{t-1}-2^{t+1}n}{2^t}\pi)$ where $n
= \lfloor \frac{k+2^{t-1}}{2^{t+1}} \rfloor$. Since
$k'+2^{t-1}-2^{t+1}n$ is in the range of $k'$, the table for
$sin(\frac{k'\pi}{2^t})$ contains all the possible values of
$cos(\frac{k'\pi}{2^t})$. Algorithm~\ref{alg:mysin} shows the sketch
of the range reduction steps.

\textbf{Range Reduction and Output Compensation for $\mathbf{cos(x)}$ and $\mathbf{tan(x)}$}. The
range reduction for $cos(x)$ is the same as that for $sin(x)$ in that
it reduces an input $x \in [-\infty, \infty]$ to a reduced input $x'
\in [-\frac{\pi}{2^{t+1}}, \frac{\pi}{2^{t+1}}]$ while minimizing the
number of precomputed values needed for output compensation.

The
output compensation formula for $cos(x)$ using $k$ and $x'$ is as
follows.

\begin{align}
    \label{cos-initial}
    cos(x) =& cos\left(\frac{k\pi}{2^t} + x'\right) = cos\left(\frac{k\pi}{2^t}\right)cos(x') - sin\left(\frac{k\pi}{2^t}\right)sin(x')
\end{align}

Using the definitions of $m$ and $k'$ defined in
Equation~\ref{sin-m-k} and the periodicity $cos(x) = cos(x - 2m\pi)$
for any integral value of $m$, we can rewrite
Equation~\ref{cos-initial} as follows.

\begin{align}
  \label{cos-intermediate}
    cos(x) = cos\left(\frac{k'\pi}{2^t}\right)cos(x') - sin\left(\frac{k'\pi}{2^t}\right)sin(x') 
\end{align}

Since $k' \in [0, 2^{t+1})$, we require a precomputed table of
$2^{t+2}$ values in total ($2^{t+1}$ for $sin(\frac{k'\pi}{2^t})$ and
$2^{t+1}$ for $cos(\frac{k'\pi}{2^t})$). Similar to $sin$, we further
reduce the size of the precomputed table to $2^{t+1}$ entries by
leveraging the property $cos(x) = sin(x+\frac{\pi}{2})$. As the
range of $k'$ is the same for both $sin(x)$ and $cos(x)$, we use a
single table containing all the possible values of
$sin(\frac{k'\pi}{2})$ for the output compensation of both
functions. In summary, our range reduction strategy for $cos(x)$ computes for
each $x$ an appropriate $k'$ and $x'$, which reduces the original
problem to approximating $sin(x')$ and $cos(x')$ for $x' \in
[-\frac{\pi}{2^{t+1}}, \frac{\pi}{2^{t+1}}]$.  Given that $tan(x) = \frac{sin(x)}{cos(x)}$, we can also implement $tan(x)$ using
the range reduction and output computation strategies described thus far and the precomputed values used for $sin(x)$ and $cos(x)$.

\setcounter{figure}{0}
\begin{figure}
  \renewcommand{\figurename}{Algorithm}%
\begin{small}  
\begin{subfigure}[t]{0.46\textwidth}
\begin{algorithm}[H]
  \SetKwProg{generate}{Function \emph{range\_reduction(x)}}{}{end}
  \generate{}{    
    $k = [ \frac{256x}{\pi} ] $\;
    $r = \frac{256x}{\pi} - k$\;
    $x' = \frac{\pi}{256}r$ \;
    $m = \lfloor \frac{k}{512} \rfloor$\;
    $k' = k - 512m $\;
    \Return $(x', k')$
  }
\end{algorithm}
\caption{}
\label{alg:mysin}
\end{subfigure}
\begin{subfigure}[t]{0.46\textwidth}
\begin{algorithm}[H]
  \SetKwProg{generate}{Function \emph{fp\_range\_reduction\_small(x)}}{}{end}
  \generate{}{
    $p0 = 0x1.45f306cp$+6$*x$\;
    $p1 = 0x1.c9c882a53f84f$p-22$*x$\;
    $p = p0 + p1$\;
    $p\_int = round(p)$\;
    $k = (int64\_t) p\_int$\;
    $r = (p0 - p\_int) + p1$\;   
    $xp = r * PI\_OVER\_256$\;
    $kp = k\  \&\  $0x1ff\;
    \Return{$(xp, kp)$}\;
  }
\end{algorithm}
\caption{}
\label{alg:sinfp-small}
\end{subfigure}
\end{small}
\caption{\small (a) High level range reduction producing a reduced
  input $x' \in [-\frac{\pi}{512}, \frac{\pi}{512}]$ and the index
  $k'$ for pre-computed tables.  Here,
  $[\frac{256x}{\pi}]$ computes the integer nearest to
  $\frac{256x}{\pi}$. (b) A FP-based implementation of Algorithm
  ~\ref{alg:mysin} that produces reduced inputs
  for small inputs (\ie, $|x| < 2^{30}$). The values $xp$ and $kp$
  denote $x'$ and $k'$ respectively.}
\end{figure}

\subsection{\textbf{Efficient Range Reduction with FP Operations}}
\label{sec:rangereduction:fp}

The range reduction described above requires computing $k =
[\frac{256x}{\pi}]$ and $x' = \frac{\pi}{256}(\frac{256x}{\pi} - k) =
\frac{\pi}{256}r$ accurately. For the inputs with relatively small
absolute values (\ie, $|x| \le \frac{\pi}{128}$), direct polynomial
approximation is possible and thus range reduction is not necessary.
Range reduction, however, is required for the remaining inputs to
obtain an efficient polynomial approximation. Implementing Algorithm
~\ref{alg:mysin} for these inputs requires computing
$\frac{256x}{\pi}$ with low latency while maintaining a large number
of bits of $\frac{256}{\pi}$ for accuracy.  The number of bits
required for $\frac{256}{\pi}$ depends on the precision of the target
representation, the exponent of the input $x$, and the number of
accurate bits needed in the reduced input $x'$. Worst case
analysis~\cite{payne-haneck:radian-reduction:signum:1983,Muller:elemfunc:book:2016,
  Ng92argumentreduction} for range reduction suggests that
approximately 200-bits of $\frac{256}{\pi}$ are sufficient for
performing range reduction with high accuracy for all 32-bit inputs.

Algorithm~\ref{alg:sinfp-small} provides an implementation of
Algorithm~\ref{alg:mysin} for small inputs. For relatively small
inputs (\ie $|x| < 2^{30}$), we observe that approximately 80-bits of
$\frac{256}{\pi}$ spread across two double-precision values suffice
for computing a reduced input with desirable accuracy.  The double  0x$\mathit{1.45f306c}$p+6 (line 2) represents the first 28-bits of
$\frac{256}{\pi}$.   The value 0x$\mathit{1.c9c882a53f84f}$p-22 represents the subsequent 53-bits of
$\frac{256}{\pi}$ obtained via round-to-nearest. Because 0x$\mathit{1.45f306c}$p+6 contains only 28-bits of
precision, the product $p0=$0x$\mathit{1.45f306cp}$+6$*x$ has at most 52 precision bits
and is thus exactly representable as a double value. By avoiding intermediate rounding errors, Algorithm
~\ref{alg:sinfp-small} can compute the integer $k$ and an accurate approximation
of the fractional value $r=\frac{256x}{\pi} - k$ using the partial
products $p0=$0x$\mathit{1.45f306c}$p+6$*x$ and $p1 =$
0x$\mathit{1.c9c882a53f84f}$p-22$*x$.

\setcounter{figure}{4}
\begin{figure}[th]
%\begin{wrapfigure}{r}{0.75\textwidth}    
  \centering{\includegraphics[width=0.95\linewidth]{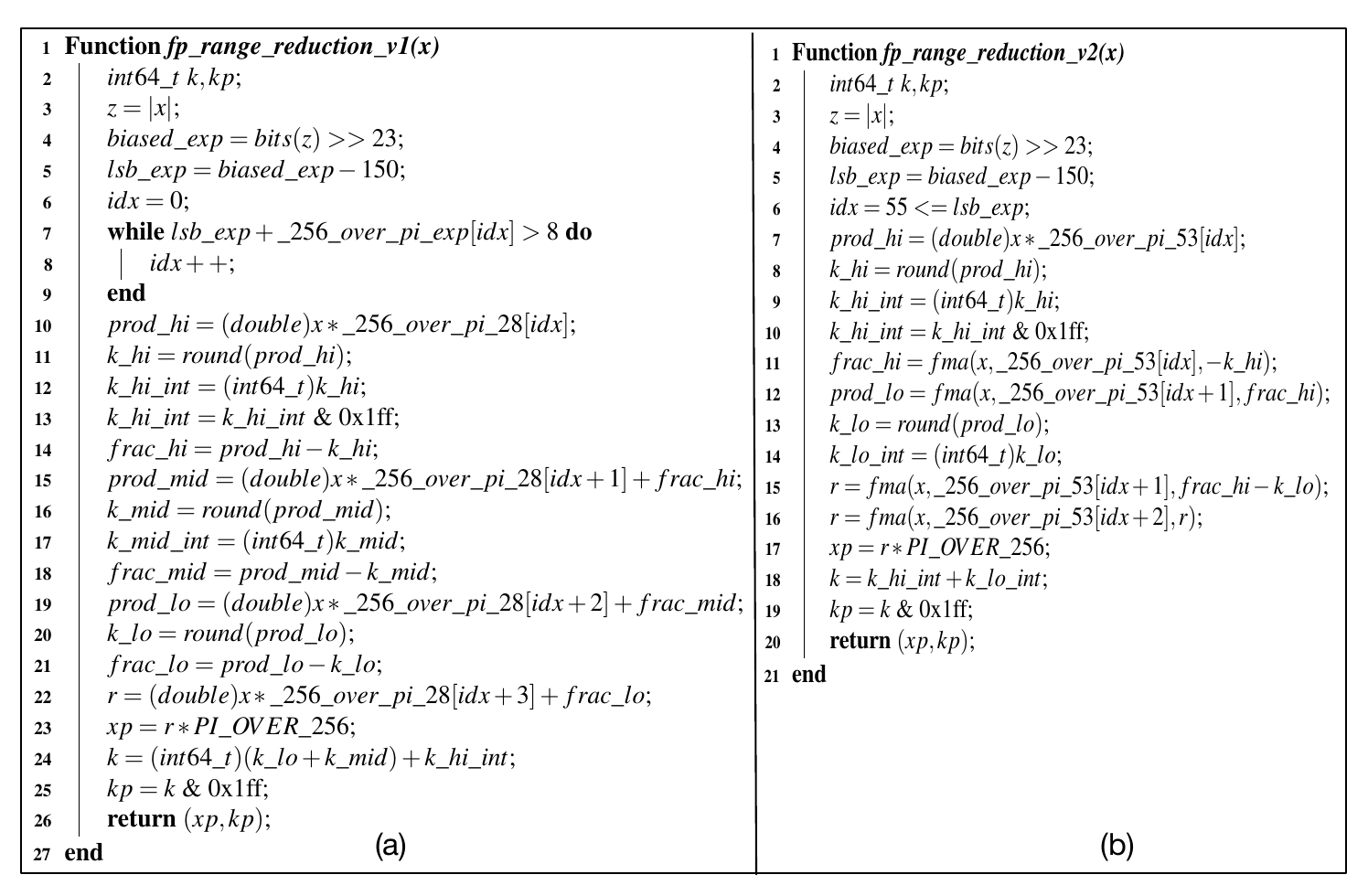}}
  \caption{\small (a) The FP-based implementation of
    Algorithm~\ref{alg:mysin} for large inputs.  We denote the reduced
    input and the lower 9-bits of $k$ as $xp$ and $kp$, respectively.
    We store 7 28-bit pieces of $\frac{256}{\pi}$ in a double array
    named $\_256\_over\_pi\_28$.  We maintain the exponent of the last
    bit of each piece in the array $\_256\_over\_pi\_28\_exp$.  Here,
    $k$ and $k\_hi\_int$ are signed 64-bit integers, $idx$ is a signed
    32-bit integer, and the rest are FP double values.  The function
    $bits(z)$ provides the IEEE-754 FP bit-pattern of the input as an
    integer. (b) An alternative FP-based strategy for large inputs.
    We store 4 53-bit pieces of $\frac{256}{\pi}$ in a double array
    named $\_256\_over\_pi\_53$. The fused-multiply-add~($fma$)
    operation that performs both FP multiplication and addition with a
    single instance of rounding.}
  \label{fig:fp-alg}
\end{figure}
%\end{wrapfigure}    

The algorithm in Figure~\ref{fig:fp-alg}(a) sketches an FP-based
implementation of the target range reduction for larger inputs (\ie,
$2^{30} \le |x|$). The algorithm's objective is to
appropriately divide the bits of $\frac{256}{\pi}$ to avoid rounding
(and the concomitant rounding errors) wherever possible.  For large
inputs, the initial bits of $\frac{256}{\pi}$ are
insufficient.  To account for a wider range of inputs, we generate
196-bits of $\frac{256}{\pi}$ using the MPFR library and store them as
28-bit pieces in an array of doubles (\ie, $\_256\_over\_pi\_28$ in
Figure~\ref{fig:fp-alg}(a)).  Each element is generated using the
round-to-zero mode (\ie, truncation).  We split $\frac{256}{\pi}$
into 28-bit pieces to ensure that no partial product
(\ie, $\_256\_over\_pi\_28[idx] * x$) incurs rounding error.  The product of a 28-bit piece of $\frac{256}{\pi}$ and a 32-bit
input with 24-bits of precision is exactly representable as a double.

The initial task in the algorithm in Figure~\ref{fig:fp-alg}(a) is to
identify the pieces of $\frac{256}{\pi}$ necessary for computing the
fractional bits of $\frac{256x}{\pi}$ and the relevant portions of
$k$, which we store as a 64-bit integer.  For a very large input, $k$ can exceed the dynamic range of \texttt{int64\_t}.
However, we only require the lower 9-bits of $k$ (see Equation~6).
Hence, we can skip over the pieces of $\frac{256}{\pi}$ for which the
exponent of the least significant bit exceeds 8 when summed with the exponent of
the least significant digit of the input (see 7-9 in
Figure~\ref{fig:fp-alg}(a)).  Once the first piece that contributes to
the least 9-bits of $k$ is identified, we proceed to compute the
partial products of $\frac{256x}{\pi}$.  At most three partial
products can contribute to the lower order bits of $k$~(lines 11-20).
After obtaining the relevant integer bits, we compute the fractional
portion of $\frac{256x}{\pi}$ denoted $r$ by accumulating the residual
fractional bits and the final partial product (\ie
$\_256\_over\_pi\_28[idx+3]*x$ in line 22).  Multiplying $r$ from this
algorithm with $\frac{\pi}{256}$, which is maintained in double
precision, produces the reduced input $x'$.

The iterative search for the first relevant piece of
$\frac{256}{\pi}$ (lines 7-9) is a key source of overhead in the
algorithm in Figure~\ref{fig:fp-alg}(a).  Another drawback is that the
limited precision of each piece in $\_256\_over\_pi\_28$ requires
computing the relevant bits of $k$ across many elements (up to 3) of
the array. We present an alternative strategy that addresses these issues in
Figure~\ref{fig:fp-alg}(b). This alternative range reduction method
splits $\frac{256}{\pi}$ into double values with 53-bits of precision
stored in the array $\_256\_over\_pi\_53$. As before, we generate each
element of $\_256\_over\_pi\_53$ using the MPFR library, albeit in the
round-to-nearest mode. The increased bit-width confines the first
piece contributing to the relevant bits for $k$ to the first two
elements.  The largest possible value of $lsb\_exp$, which represents
the exponent of the least significant bit of the input, is $254-150 =
104$ given that the largest possible unbiased exponent for a
non-infinity, non-$NaN$ 32-bit float is 254.  The exponent of the
least significant bit of $\_256\_over\_pi\_53[1]$ is
$-99$, and thus the largest possible exponent for the least significant bit of
the product $\_256\_over\_pi\_53[1]*x$ is $104-99=5$. Therefore,
either the first or second element of $\_256\_over\_pi\_53$ is
necessary to compute the 9 lowest bits of $k$ for all inputs.  Whether
the first element is needed depends on the value of $lsb\_exp$. The
exponent of the least significant bit of $\_256\_over\_pi\_53[0]$ is
$-46$. For all inputs $x$ such that $lsb\_exp \ge 55$, the exponent of
the least significant bit of $\_256\_over\_pi\_53[0]*x$ evaluated in
infinite precision is at least $-46+55 = 9$.  As such, the first piece
of $\_256\_over\_pi\_53$ does not contribute to the 9 lowest bits $k$
for any input such that $lsb\_exp \ge 55$. The comparison $55 \le
lsb\_exp$ (line 6) thus determines the first relevant piece of
$\_256\_over\_pi\_53$.  By limiting the starting point to the first
two elements of $\_256\_over\_pi\_53$, the algorithm in
Figure~\ref{fig:fp-alg}(b) avoids the iterative search in
Figure~\ref{fig:fp-alg}(a).

Unlike the algorithm in Figure~\ref{fig:fp-alg}(a), which divides
$\frac{256}{\pi}$ into 28-bit pieces to avoid rounding during
multiplication, the algorithm in Figure~\ref{fig:fp-alg}(b) is
susceptible to more intermediate rounding errors as the product of $x$
and an element of $\_256\_over\_pi\_53$ may not be exactly
representable as a double.  To minimize this rounding error, we use
fused-multiply-add instructions (lines 11-16 in
Figure~\ref{fig:fp-alg}(b)), which perform multiplication and addition
sequentially with a single instance of
rounding~\cite{Boldo:reduction:toc:2009}. By computing the partial
products (\ie, $\_256\_over\_pi\_53[idx]*x$) while subtracting
portions of the previous products that contribute to integer bits
through \texttt{fma} instructions (see lines 11 and 15), we are able
to compute the fractional bits composing $r$ (line 16) with sufficient
precision in Figure~\ref{fig:fp-alg}(b).

\subsection{\textbf{Efficient Range Reduction with Integer Operations}}
\label{sec:rangereduction:integer}

After we implemented the above range reduction strategy with FP
operations, we observed that the resulting implementations produced
correctly rounded results but were slower than mainstream
libraries. Hence, we subsequently explored integer-based
implementations that compute $\frac{256x}{\pi}$. Moreover, we can compute with
more precision than FP doubles by using integers
(\ie, \texttt{uint64\_t} and \texttt{uint128\_t}), which can reduce the
number of intermediate products.
In the earlier FP-based strategies, maintaining
the pieces of $\frac{256}{\pi}$ and the partial products as doubles
helped us identify the integer $k$ and the fractional parts for the
reduced input easily.  With our integer based implementation, we
maintain the pieces of $\frac{256}{\pi}$ and partial products as
integers and we track the exponent implicitly. Finally, we compute the
reduced input and $k$ using bitwise shift operations.

%\begin{figure}[htb!]
\begin{wrapfigure}{r}{0.7\textwidth}  
  \centering{\includegraphics[width=0.99\linewidth]{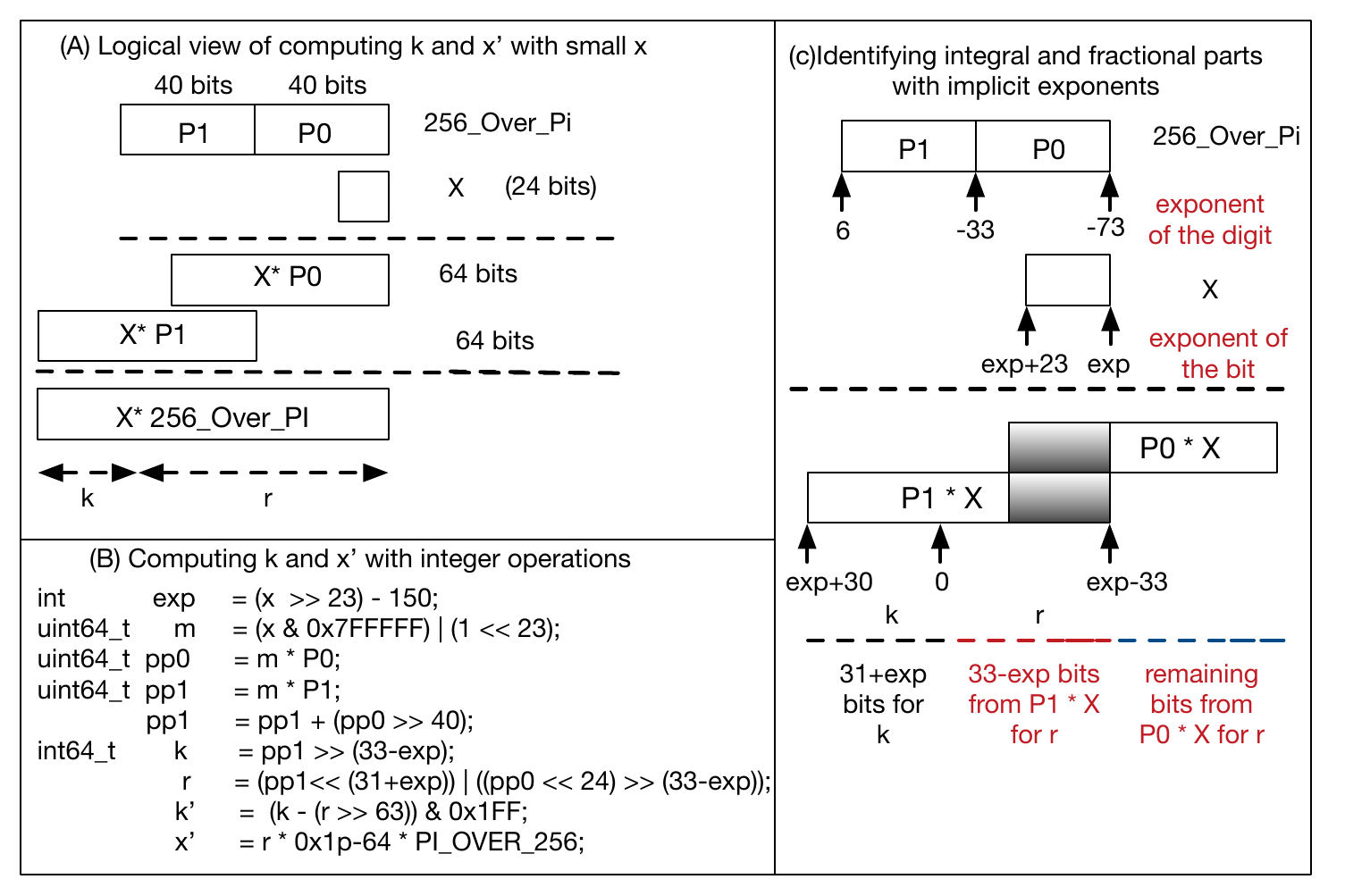}}
  \caption{\small Range reduction for small inputs when $\frac{\pi}{128} \leq |x| <
    2^{30}$ and $t=8$. Here, we are creating a reduced input $x' \in
    [-\frac{\pi}{512}, \frac{\pi}{512})$. (A) Computation with 80 bits
      of x * 256\_Over\_Pi with 40 bits each in two 64-bit integers P1
      and P0. (B) Computing $k$ and $r$ with integer and bitwise
      operations. (c) Intuition for finding the bit whose exponent is
      0 when we perform computation with integers and do not
      explicitly maintain exponents.}
    \label{fig:pi-small}
\end{wrapfigure}    
%\end{figure}

\textbf{Range reduction with small inputs.} For inputs such that $|x| < 2^{30}$ (excluding special case inputs
that do not require range reduction), we compute $k$ and the reduced
input using the first 80-bits of
$\frac{256}{\pi}$.  As we want to primarily compute with integers, we
represent the significand of the input $x$ as an integer (\ie,
multiplying by $2^{23}$, which is equivalent to subtracting 23 from
the exponent of $x$).  For these inputs, the integer $k$ is contained in the first 64-bits of the product $\frac{256x}{\pi}$
and we can generate precise reduced inputs using
80-bits of $\frac{256}{\pi}$. We generate 80 bits of $\frac{256}{\pi}$
and store 40 bits of $\frac{256}{\pi}$ in two 64-bit numbers $P1$ and
$P0$. We store 40-bits of $\frac{256}{\pi}$ each in $P1$ and $P0$
because we can store $P1*x$ and $P0*x$ exactly as 64-bit integers.
The partial products are added as shown in
Figure~\ref{fig:pi-small}.

The final task is to identify the position
of the binary point in the resultant product, which will help us
identify $k$ and the bits belonging to the reduced input $x'$.
Figure~\ref{fig:pi-small}(c) pictorially provides the intuition for finding the position of the binary point.
The exponent of the most significant bit of $\frac{256}{\pi}$ in $P1$ is
$6$. The exponent of the least significant bit in $P1$ is
$-33$. When we compute $P1*x$, the exponent of the
least significant bit of this partial product is $exp-33$, where $exp$ represents the exponent of the least significant bit of the original input. 
Since the partial product can have up to 64 precision bits, the exponent of the most significant bit
would be at most $exp-33+63=exp+30$. The binary point (\ie, exponent $0$)
lies within this partial product (\ie, between the exponents $exp+30$ and
$exp-33$).  We can compute the integral part of the product (\ie, $k$)
with a right shift operation by $33-exp$ because there are $0-(exp-33)
= 33-exp $ bits after exponent $0$.
Next, we accumulate all bits past the binary point in the products
$P1*x$ and $P0*x$ into a 64-bit integer $r$.  We first need to place the $33-exp$ fractional bits 
in the partial product  $P1*x$ as the most significant bits
of $r$. We accomplish this with a left shift by $31+exp$ because
there are $64-(33-exp)$ non-fractional bits in $P1*x$ pertaining to the integer portion
$k$. Then, we need to collect the remaining bits for the reduced input from $P0*x$. 
The leading 24-bits of $P0*x$, which are added to the partial product $P1*x$, 
can be removed with a left shift by 24. We subsequently place the most significant portions of the remaining bits of
 $P0*x$ (\ie, $(P0*x)<<24$) after the $33-exp$ initial bits in $r$ obtained from $P1*x$. 
This can be accomplished by performing a right shift on $(P0*x)<<24$ with a shift amount of $33-exp$. Finally, we compute the
reduced input by multiplying $r$ with the floating point value
$2^{-64}$ and $\frac{\pi}{256}$ as shown in
Figure~\ref{fig:pi-small}(B).

\begin{figure}
  \centering{\includegraphics[width=0.9\linewidth]{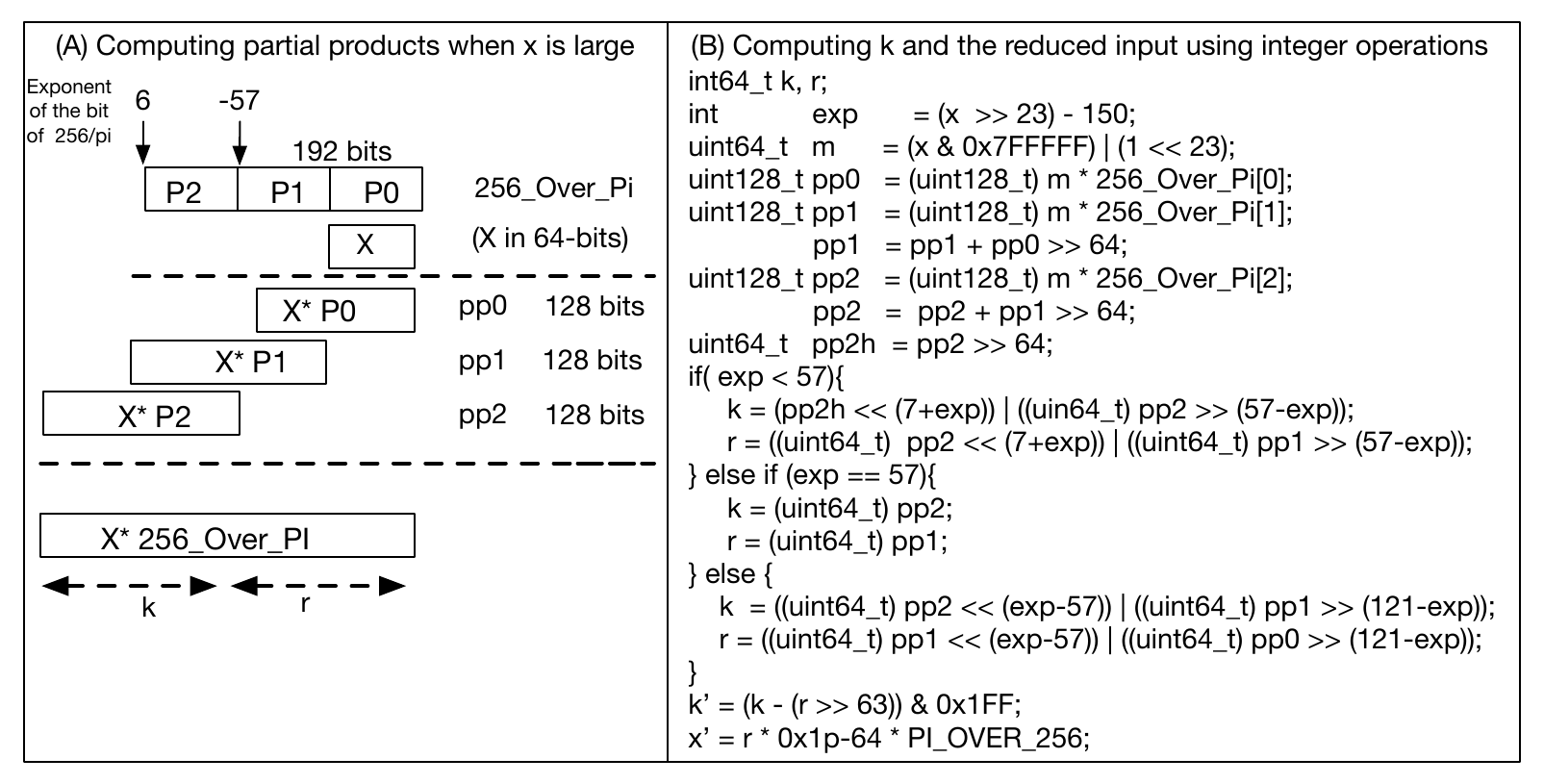}}
  \caption{\small Range reduction for large inputs when $2^{30} \leq
    |x|$ when $t=8$ to produce the reduced input $x' \in
    [-\frac{\pi}{512}, \frac{\pi}{512})$. (a) We show the logical
      computation of $\frac{256x}{\pi}$. (b) The computation with
      integer and bitwise operations. Each piece of 256\_Over\_Pi
      (\ie, P2, P1,and P0) is 64-bits.}
    \label{fig:pi-large1}
 \end{figure}

\textbf{Range reduction with large inputs}. When the input
$|x|>2^{30}$, we use 192 bits of $\frac{256}{\pi}$ maintained as three
64-bit integers. We also maintain the 24 bits in the significand of
input $x$ as a 64-bit integer.  We compute the partial products
between $x$ and 64-bit pieces of $\frac{256}{\pi}$ and represent them
as 128-bit integers as shown in Figure~\ref{fig:pi-large1}(A).  After
computing the partial products, we identify the position of the binary
point based on the exponent of $x$ and the exponents of each piece of
$\frac{256}{\pi}$.
Figure~\ref{fig:pi-large1}(B) provides the bitwise
operations that identify $k$ and fractional bits required for the
reduced input. The exponent of the most significant bit of
$\frac{256}{\pi}$ is 6 and the the exponent of the least significant
bit of P2, the first piece of $\frac{256}{\pi}$, is -57.  When the
exponent of the least significant bit of $x$, represented by $exp$, is
less than 57, the integer bits pertaining to $k$ will be completely
contained in the partial product \texttt{pp2}.
The remaining step involves extracting 64 fractional bits for the
reduced input through bitwise shifts akin to those performed on the
smaller inputs in Figure~\ref{fig:pi-small}(B).

%\begin{wrapfigure}{r}{0.7\textwidth}
%\end{wrapfigure}        
%\end{figure}
%% \begin{figure}
%%   \centering
%%   \begin{subfigure}{0.5\textwidth}
%%   \includegraphics[width=0.9\linewidth] {figures/rr-large1.pdf}
%%   \caption{}
%%     \label{fig:pi-large1}
%% \end{subfigure}
%% %
%% \begin{subfigure}{0.5\textwidth}
%%   \includegraphics[width=0.9\linewidth]{figures/rr-large2.pdf}
%%   \caption{}
%%     \label{fig:pi-large2}
%% \end{subfigure}
%% \caption{Range reduction for large inputs when $x \geq 2^{21}$ when
%%   $t=8$. Here, we are creating a reduced input $x' \in [0,
%%     \frac{\pi}{256})$. (a) We show the logical computation of
%%     $\frac{256}{\pi}x$. (b) We show the computation with integer and
%%     bitwise operations. Each piece of 256\_Over\_Pi (\ie, P3, P2,
%%     P1,and P0) is 64-bits.}

%% \end{figure}

When $exp$ is equal to 57, there is no need to perform any shift
operations. Here, the relevant bits for $k$ are the lower 64-bits of
the partial product \texttt{pp2} and the fraction bits for the reduced
input are the lower 64-bits of partial product
\texttt{pp1}.  When $exp$ is greater than $57$, $k$
will be computed using the lower 64-bits of the partial product
\texttt{pp2} and portions of \texttt{pp1}. The reduced input will be computed
using the lower 64-bits of the partial product \texttt{pp1} and portions of
\texttt{pp0}.
Everything is computed with integers prior to multiplying
$r$ with $2^{-64}$ and $\frac{\pi}{256}$ to generate the reduced input
as shown in Figure~\ref{fig:pi-large1}(B). All shift amounts in
Figure~\ref{fig:pi-large1}(B) are positive and less than 64, which avoids
undefined behavior. The largest value of $exp$ in
Figure~\ref{fig:pi-large1}(B), which represents the exponent of the least
significant bit of the input, is 104.  This is because the largest
exponent of any 32-bit FP number is 127.

\iffalse
In summary, this alternative
range reduction strategy using integer and bitwise operations enables
us to generate fast and correct trigonometric functions.
\fi

\section{Experimental Evaluation}
We have developed various implementations of trigonometric functions
that produce correctly rounded results for all inputs for all FP
representations up to 32-bits. Our prototype uses the MPFR
library~\cite{Fousse:toms:2007:mpfr} as the oracle to generate the
library. We developed new range reduction techniques, inference
techniques for reduced intervals, and new polynomials for the
trigonometric functions with the \rlibm approach. To evaluate our
functions for correctness and performance, we compare it against
mainstream libraries (\eg, GLIBC's float and double libm) and the
correctly rounded Core-Math
library~\cite{sibidanov:core-math:arith:2022}.
We conducted our experiments on a $2.10$GHz Intel Xeon(R) Silver 4310
server with $256$GB of RAM running Ubuntu 24.04.1 LTS and used performance counters to measure the time taken.

\textbf{Ability to produce correct results.}
Our functions produce correctly rounded results for all
representations from 10-bits to 32-bits for all inputs, which we
tested with complete enumeration similar to bounded model
checking.
\iffalse
Our functions produce the correctly rounded result for the
34-bit FP representation with the round-to-odd mode, which makes
double-rounding innocuous.
\fi
In contrast, GLIBC's float libm does not
produce correctly rounded results for 32-bit float inputs for
$sin$, $cos$, and $tan$ even for one representation and has several thousand
incorrect results.  GLIBC's double libm is more accurate than GLIBC's
float libm but still produces wrong results for some inputs.
For the 32-bit float representation, both our functions and Core-Math
produce correctly rounded results for all inputs. However, Core-Math
does not produce correctly rounded results for all inputs from 10-bits
to 32-bits, which is due to double-rounding errors.

\textbf{Performance due to our range reduction optimizations}.
For the FP-based approach, we developed two implementations for
$sin$, $cos$, and $tan$ that apply the range reduction algorithms shown in Figure~\ref{fig:fp-alg}(a) and
Figure~\ref{fig:fp-alg}(b).  We have also developed implementations
for these functions that use the integer-based range reduction
depicted in Figures~\ref{fig:pi-small} and
~\ref{fig:pi-large1}. Lastly, we experimented with applying different
approaches for different sub-domains. Specifically, we implemented
versions of $sin$, $cos$, and $tan$ that use the FP-based approach for smaller
inputs (see Algorithm~\ref{alg:sinfp-small}) and an integer-based
approach for larger inputs (see Figure~\ref{fig:pi-large1}).
%

%\begin{wrapfigure}{r}{0.75\textwidth}    
\begin{figure}[t]
\centering
\begin{subfigure}{0.48\textwidth}
\centering
\includegraphics[width=\linewidth]{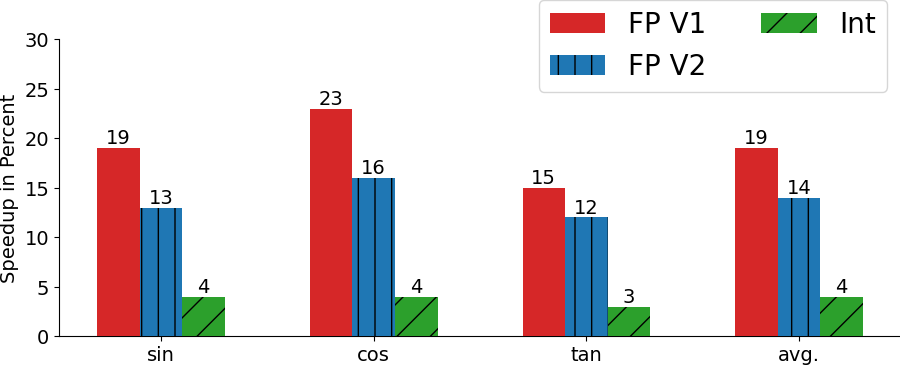}
\caption{}
\label{fig:speedup-fp-int}
\end{subfigure}
\begin{subfigure}{0.48\textwidth}
\centering
\includegraphics[width=\linewidth]{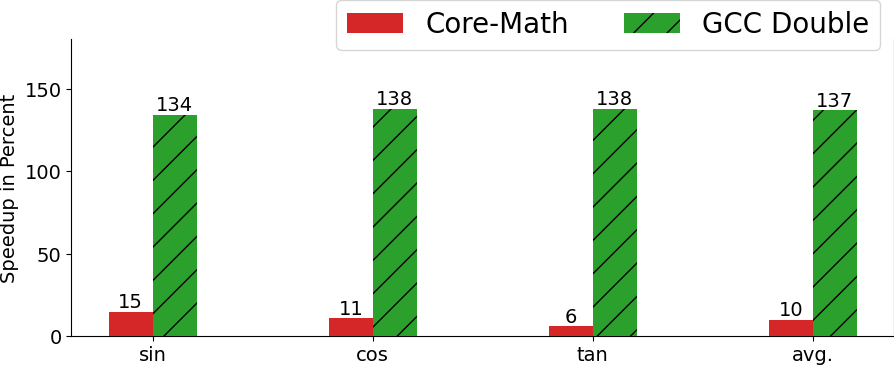}
\caption{}
\label{fig:speedup-other}
\end{subfigure}
\caption{\small (a) Performance improvement using our approach that
  uses FP-based range reduction for small inputs and integer-based
  range reduction for larger inputs over other approaches. (b)
  Performance improvement of the fastest \rlibm implementation over
  other libraries.}
\label{fig:performance}
%\end{wrapfigure}
\end{figure}

Among these strategies, applying a FP-based approach for smaller
inputs and an integer-based approach for larger inputs leads to the
best performance for all three functions. Figure~\ref{fig:speedup-fp-int} reports the speedup of this
approach over the FP-based approach detailed in
Figure~\ref{fig:fp-alg}(a) (\textbf{FP V1}), the alternative FP-based
approach in Figure~\ref{fig:fp-alg}(b)(\textbf{FP V2}), and the
integer-based approach (\textbf{Int}). On average, our most efficient,
hybrid range reduction achieves a 19\% speedup against the
initial FP-based approach. This result highlights the advantage of
employing an integer-based range reduction for large inputs. We
attribute this speedup to the larger amount of precision relative to
64-bit doubles (\ie 53-bits of precision) available with integer
representations (\ie, \texttt{uint64\_t} and \texttt{uint128\_t}). The
larger amount of precision available for each piece of
$\frac{256}{\pi}$ and intermediate outputs in the integer-based
approach reduces the number of partial products required to obtain a
reduced input with sufficient accuracy. Moreover, the bit-wise
operations employed for the integer-based approach greatly simplifies
identifying the portions of $\frac{256x}{\pi}$ relevant to the reduced
input and the lower order bits of $k$ for any given input.
Alternatively, the speedup of the hybrid approach over the
integer-based approach (4\% on average) indicates that FP operations
provide a more ideal solution when the inputs have low magnitudes and
the relevant portions of $\frac{256}{\pi}$ are confined to the initial
set of bits. We attribute this result to the observation that the
integer portion of $\frac{256x}{\pi}$ for small inputs (\ie, $|x| <
2^{30}$) are small enough to be exactly representable using a single
double-precision value. For such cases, the results indicate that
computing the first few partial products of $\frac{256x}{\pi}$ and
subtracting away integer bits identified through FP rounding
operations are sufficient for obtaining reduced inputs with the
desirable level of accuracy with low overhead.

Figure~\ref{fig:speedup-other} reports the performance speedup of our
fastest implementations when compared to other libraries.  On average,
our functions are 10\% and 137\% faster than Core-Math's functions and
GLIBC's double functions respectively.  The performance speedups over
Core-Math are smaller because Core-Math's implementations are also
well optimized with a range reduction strategy appropriately mixing
both FP and integer-based approaches. Unlike Core-Math's functions,
which produce correct results only for 32-bits, our functions produce
correct results for multiple representations from 10-bits to 32-bits
and all five rounding modes in the IEEE standard.

\vspace{-0.15in}
\section{Conclusion and Future Work}
We extend the \rlibm approach to trigonometric functions by paying
careful attention to the amount of precision required in handling
$\frac{256}{\pi}$, while obtaining performance using integer and
bitwise operations. We have been collaborating with the developers of
mainstream math libraries to incorporate these methods, which will
enable push-button usage. We are also participating in the standards
committees to mandate correct rounding in the next version of the
IEEE-754 standard. Finally, we want to explore correctly rounded
libraries for GPU platforms in the future.

\vspace{-0.15in}
\section{Acknowledgments}
  We thank the VSS reviewers and Bill Zorn for their feedback. This
  material is based upon work supported in part by the research gifts
  from the Intel Corporation and by the National Science Foundation
  with grants: 2110861, 2312220, and 1908798.

\newpage

\bibliographystyle{eptcs}
\newpage
\bibliography{references.bib}
\end{document}